\documentclass[twocolumn,10pt]{article}

\usepackage[margin=1in]{geometry}
\usepackage[medium,compact]{titlesec}
\usepackage{alltt}

\usepackage{listings}
\newif\ifpdf
\ifx\pdfoutput\undefined
  \pdffalse
  \pdfoutput=0
\else
  \ifx\pdfoutput\relax
    \pdffalse
    \pdfoutput=0
  \else
    \ifnum\pdfoutput=0
      \pdffalse
    \else
      \pdftrue
    \fi
  \fi
\fi

\ifpdf      \usepackage{hyperref}
      \usepackage[pdftex]{graphics}
      \usepackage{epstopdf}
\else\usepackage{graphics}
     \usepackage[hyphens]{url}
     \usepackage{cite}
\fi

\usepackage[T1]{fontenc}
\usepackage{times}
\usepackage{textcomp}
\usepackage{subfigure}
\usepackage{amsmath,amsthm,amsfonts,amssymb}
\usepackage{epsfig}
\usepackage{cite}
\usepackage{color}
\usepackage{xspace}

\newcommand{\stitle}[1]{\vspace{1pt}{\noindent\bf #1:}}

\newcommand{\eat}[1]{}

\newcommand{\abs}[1]{\lvert#1\rvert}

\newcommand{\vecx}{\mathbf{x}}

\def\E{{\rm E}}

\long\def\comment#1{}

\newcommand{\R}{\ensuremath{\mathbb R}}

\title{
\vspace{-1.2em}
\bf\Large Mantis: Predicting System Performance through Program Analysis and Modeling
\vspace{-0.7em}
}

\author{Byung-Gon Chun$^\dagger$, Ling Huang$^\dagger$, Sangmin Lee$^\star$, Petros Maniatis$^\dagger$, Mayur Naik$^\dagger$\\
        \textit{$^\dagger$Intel Labs Berkeley, $^\star$University of Texas at Austin}}

\date{}

\renewenvironment{thebibliography}[1]{%
    \begin{oldthebibliography}{#1}%
    \setlength{\parskip}{0ex}%
    \setlength{\itemsep}{0ex}%
}%
{%
    \end{oldthebibliography}%
}

\begin{document}

\maketitle

\begin{abstract}
We present \emph{Mantis}, a new framework that automatically predicts program performance with high accuracy. Mantis integrates techniques from programming language and machine learning for performance modeling, and is a radical departure from traditional approaches. Mantis extracts program features, which are information about program execution runs, through program instrumentation. It uses machine learning techniques to select features relevant to performance and creates prediction models as a function of the selected features. Through program analysis, it then generates compact code slices that compute these feature values for prediction. Our evaluation shows that Mantis can achieve more than 93\% accuracy with less than 10\% training data set, which is a significant improvement over models that are oblivious to program features. The system generates code slices that are cheap to compute feature values.
\end{abstract}

\section{Introduction}

Today's programs are numerous and become more and more complex. For example, services running in data centers are 
often large scale, and perform complicated operations depending on input workload. Predicting how applications
will behave for given input workload is key to helping users and operators better manage those applications.

Predicting metrics (e.g., performance, resource consumption) has great applications in many usage scenarios. First, prediction of execution time of a service request can be used for better workload management~\cite{Ganapathi2009}. If the request is likely to violate service level agreements, the system can drop the request and allocate resources to other requests. Second, in scheduling applications such as MapReduce~\cite{mapreduce, Quincy}, if we can predict execution time of tasks, we can then schedule jobs more optimally by considering where to map individual tasks to candidate resources and perform speculative execution in a timely fashion without spawning unnecessary processes. Third, prediction can help with better resource provisioning~\cite{DynProv,Bodik2009a,Bodik2009b} (e.g., how many servers should I use to run this job? Should I add more servers?). Fourth, with prediction, we can detect performance anomaly. If an operation takes much longer than a predicted time, we label it an anomaly for troubleshooting purposes. Finally, prediction can answer what if questions --- how system behavior changes when input workload changes or system configuration changes~\cite{Li2010,Tariq2008,Chen2009}.

Despite all these opportunities and demands, prediction has not been in the mainstream. This is because it is very difficult to predict metrics with high accuracy for current practices --- analytically modeling the system or treating the system as a black box and generating a transfer function between input workload and output response. System execution inherently depends on program semantics (i.e., internals of how the program works), thus prediction depends on program semantics. For example, certain metadata of programs (e.g., image resolution and depth) is a cache of program semantics. One way to obtain program semantics is to ask its details to its developers. In reality, however, this is not feasible due to the abundance and complexity of programs. In this work, we aim to automatically extract program semantics without developers of the program and use them to create better prediction models for system performance (execution time). In particular, we focus on predicting with different input workload in the same environment (i.e., machine).

\emph{Mantis} is a system that achieves this goal by combining programming language and machine learning techniques in a novel way (Section~\ref{sec:overview}).
Mantis consists of three key components: feature instrumentation, model generation, and code snippets generation for computing feature values when predicting.
To capture program semantics without programmer assistance, we begin by extracting a potentially large number of program features 
that capture the characteristics of program execution by running programs instrumented with code analysis (Section~\ref{sec:inst}). 
Next, we use machine learning techniques to select important, relevant features to create the prediction models (Section~\ref{sec:model}). 
Finally, program slicing computes small code snippets that compute features 
needed by the model for prediction (Section~\ref{sec:slicing}). 
This component also guides model generation to choose features 
that can be computed cheaply.

\begin{figure*}[t]
\centering{\includegraphics{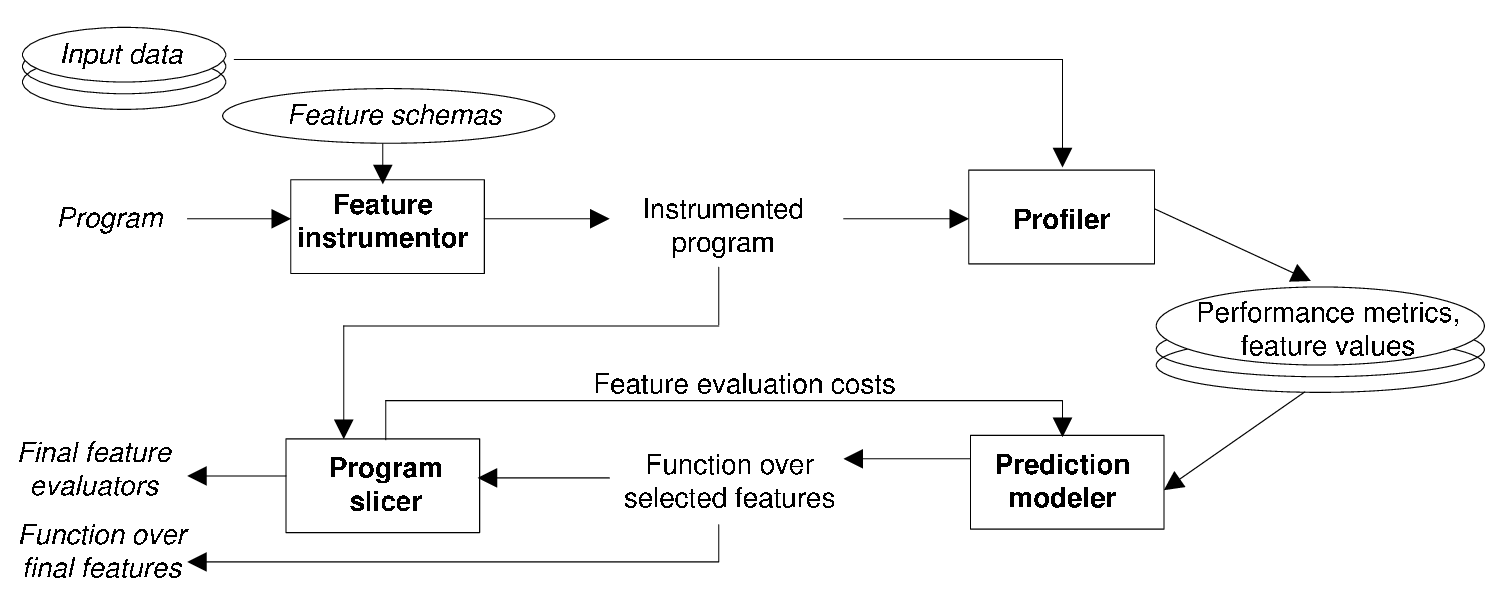}}
\caption{Mantis architecture.}
\label{fig:arch}
\end{figure*}
We evaluate our system by applying the framework to two applications, 
Lucene search engine and ImageJ, an image processing applications 
(Section~\ref{sec:eval}).
We show that Mantis can achieve more than 93\% accuracy with less than 10\% training data set. We compare the model generated by Mantis with blackbox approaches that rely on workload size and input configurations (e.g., command-line arguments) and show that Mantis can significantly outperform these models. We explain how slicing can benefit model generation and the overhead of computing selected features. Finally, in Section~\ref{sec:related} we discuss related work, and conclude with future research directions in Section~\ref{sec:conc}.

\section{Mantis Overview}
\label{sec:overview}

\subsection{Approach}
We address the problem of predicting performance metrics quickly without actually running the entire program. 
Traditionally, researchers have taken a stance in two camps for prediction --- modeling systems analytically (e.g., queuing theory), or treating the system as a black box and creating a model between input (workload) and output (performance). However, these approaches do not work well due to their inherent limitations, i.e., the lack of knowledge about the program. We take a new white-box approach to generating prediction models. Unlike traditional approaches, simply put, we extract information from execution of the program that contains a plethora of information.
In particular, we extract as many features as possible 
from programs for given input data if extracting features 
incurs little overhead, and rely on machine learning techniques
to process the large amount of information dumped out. 
Machine learning techniques can infer 
key features from voluminous information and construct a robust model 
that predicts performance based on new program features. 
In summary, our approach solves the prediction system problems by 
combining programming language and machine learning techniques 
in a novel way.

To achieve our goal,
we need to address three key questions:
\begin{enumerate}
\item What are good program features? How do we extract these feature values?
\item Among many features, which ones are relevant to performance metrics? 
How do we model performance with relevant features?
\item How do we automatically generate code to compute 
feature values for prediction?
\end{enumerate}

We present Mantis, a new prediction architecture that addresses 
the three questions above. There are three main components, 
each of which addresses a key question.

\subsection{Architecture}

Figure~\ref{fig:arch} shows the Mantis architecture, a novel prediction framework that combines programming language techniques with machine learning techniques. This architecture shows the offline part for generating prediction models.

Mantis consists of three major components: feature instrumentation and profiling, prediction model generation, and feature evaluator generation. The feature instrumentor \emph{analyzes the code} of the program and automatically adds instrumentation code that extracts program features. Then the profiler runs this instrumented program with sample input data to collect performance metrics and feature values. This profiling can generate a large number of features within the budget of instrumentation overhead. Then, the model generator runs \emph{machine learning} algorithms to generate a prediction model, i.e., select a subset of key features that are relevant to the performance metrics and create a function of the selected features to predict the performance metrics with high accuracy. To use the model, we need a way to compute feature values. The feature evaluator generator uses \emph{program slicing} to automatically extract small code snippets (which we call feature evaluators) that compute feature values from the instrumented program. Ideally, for a feature evaluator, the technique includes only program statements that affect the feature value of the evaluator. Finally, there is a feedback loop from the slicer to the model generator. We may not be able to use some of the features selected by the model generator. If a selected feature is expensive to compute (e.g., we have to run the entire program to compute the feature value), we reject the feature by notifying the cost of computing the feature value
to the model generator.
The prediction model generator creates a new model after excluding
the rejected feature(s). This loop may run multiple times depending on scenarios. When the program slicer can generate all the feature evaluators of the selected features (cheaply), the entire process ends, and the tool produces final features, prediction model, and feature evaluators.

\begin{figure}[t]
\centering{\includegraphics{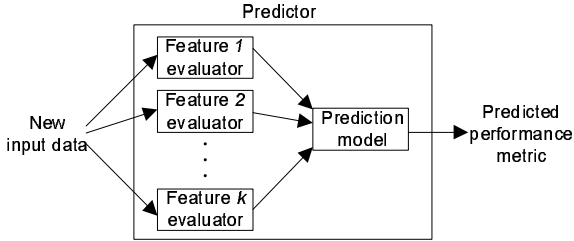}}
\vspace{-0.1in}
\caption{Online predictor.}
\label{fig:predictor}
\vspace{-0.1in}
\end{figure}

Once a prediction model is generated, it is used to forecast a
performance metric of interest for a new input as shown in
Figure~\ref{fig:predictor}. The example has $k$ feature evaluators.
The new input is sent to each feature evaluator to compute its
feature value, and the prediction model computes an estimate using all
the feature values.

In the following, we explain these components in detail
and evaluate the system.

\section{Feature Instrumentation}
\label{sec:inst}

We extract program features relevant to performance (e.g., execution time).
We choose features with the following goals in mind. First, the features
should capture the behavior of program performance. Second, the features
should be accurate and easy to compute. For example, we avoid relying
on inaccurate timer resolution. Third, the features should be collected
with low overhead. We aim to run our instrumented program to collect
feature values and performance metrics at the same time
instead of running the original program to get performance metrics,
running the instrumented program to get feature values, and joining the data.
The latter is not accurate when the program has non-determinism. 

In the following, we first describe what program features we instrument
and then present how to create instrumented programs.% with the details
We use Java programs as examples, but our techniques are generally applicable
to other programming languages.

\subsection{Features}

The features we choose are loop counts, branch counts,
and variable values in different versions. 
We discuss the rationales of choosing these features below.

\paragraph{Loops}
When a program repeats computation, the execution time depends on
how many times the program repeats.
We introduce loop counts to capture this behavior.
We instrument all loop constructs (e.g., while and for) in the program.
If there are nested loops, we add a loop count for each loop.
The following example shows a nested loop. The outer loop performs reading a
line from a file, and the inner loop performs a search operation 
$n$ times.
\begin{figure}[h]
{\small
\begin{verbatim}
// original code
while(line=readLine()) {
  for (int i=0; i<n; ++i)
    search(line, i);
}
// instrumented code
while(line=readLine()) {
  ++mantis_loop_cnt1;
  for (int i=0; i<n; ++i) {
    ++mantis_loop_cnt2;
    search(line, i);
  }
}
\end{verbatim}
}
\label{fig:example-loop}
\vspace{-0.2in}
\end{figure}

\paragraph{Method invocation} Another way to repeat computation is to use a recursive procedure. The execution time depends on how many times the program invokes recursive methods. To capture this behavior, we introduce method invocation counts, each of which is incremented when a method is invoked. 
The following example shows an example program that traverses a tree structure and computes an aggregated metric.

\begin{figure}[h]
{\small
\begin{verbatim}
// original code
process(node n) {
  if (cond) return;
  process(n.l);
  process(n.r);
  compute(n);
}
// instrumented code
process(node n) {
  ++mantis_methodinv_cnt;
  if (cond) return;
  process(n.l);
  process(n.r);
  compute(n);
}
\end{verbatim}
}
\label{fig:example-recursion}
\vspace{-0.2in}
\end{figure}

\paragraph{Branches}
Often the execution time changes depending on which control flow path
the program takes. This can be captured by adding branch information.
The following example shows that depending on the conditional, the program
takes two different paths with very different execution times.
\begin{figure}[h]
{\small
\begin{verbatim}
// original code
if (flag) { lightweightCompute(); }
else { heavyCompute(); }
// instrumented code
if (flag) {
  ++mantis_branch_cnt1;
  lightweightCompute();
} else {
  ++mantis_branch_cnt2;
  heavyCompute();
}
\end{verbatim}
}
\label{fig:example-branch}
\vspace{-0.2in}
\end{figure}

We add a branch counter for each branch in the program. It counts how many
times a particular branch is taken. For example, if the program takes
a particular branch once depending on a conditional, the counter
value is either 1 or 0. If a branch is taken multiple times, 
its value reflects that.

\paragraph{Variable values}
We also instrument versions of variable values to characterize
the program execution. We focus on primitive variables (short, int, long,
float, double, char, and boolean variables) and collect
the first $k$ values whenever a variable is assigned to a value.
Our intuition is that
often the variable values obtained from input parameters and configurations
are changing infrequently, and these values tend to affect program execution
by changing control flow. We track both class field variables and local
variables.

In the following example, the execution time of the program is dominated
by the input argument $n$ of $compute()$, which comes from a preprocessed variable, which is done quickly. We collect the assigned
value of $n$ as one of our program features.
\begin{figure}[h]
{\small
\begin{verbatim}
// original code
n = preprocess();
compute(n);
// instrumented code
n = preprocess();
mantis_n_data[cur_ptr++] = n;
compute(n);
\end{verbatim}
}
\label{fig:example-variable-value}
\vspace{-0.2in}
\end{figure}

\paragraph{Exception counts} For certain inputs, the program may take a 
control flow that throws and handles errors. This path is not a common case
the program takes, so the execution time of the program is likely to change 
significantly. We add an exception count for each exception handling part of
the program to capture this behavior. 
\begin{figure}[h]
{\small
\begin{verbatim}
// original code
try { compute(); }
catch (Exception e) { error(e); }
// instrumented code
try { compute(); }
catch (Exception e) { 
  ++mantis_ex_cnt;
  error(e); 
}
\end{verbatim}
}
\label{fig:example-exception}
\vspace{-0.2in}
\end{figure}

To collect these features in multi-threaded object-oriented programs,
we need to summarize the features across objects and across threads.
We sum up loop count and branch count across objects, and also keep
a single array of a variable for all objects created. To handle multiple
threads we maintain a separate instrumentation object that captures
features per thread, and merge feature values at the end of program
execution. For loop and branch counts, we sum up those counts across threads.
For versions of variables, we compute the mean of each version across threads.

\subsection{Instrumentor}

To instrument program to obtain program features, we perform 
code analysis and transformation. In particular, we use source code 
analysis\footnote{We can also implement our instrumentation 
using bytecode analysis. With bytecode, we do not have local variable 
names we can refer to unless source code is compiled with 
the Java compiler debug option.}
to construct abstract syntax trees (ASTs), and manipulate the 
constructed ASTs by adding new nodes representing loop counts, 
method invocation counts, branch counts, exception counts, 
and variable versions to the trees. 

We use the instrumented program to capture the execution time 
of the original program as well as to capture program
feature values. To achieve low overhead, we employ three 
techniques. First, we perform selective profiling of programs. 
We focus on application programs and do not instrument system 
libraries (e.g., toString() or equals()). Second, we use a 
procedure that removes instrumentation from the part that incurs high
overhead until the overall instrumentation overhead is below 
our threshold (e.g., 5\% of the original program execution time). 
Third, to avoid synchronization overhead of multiple threads
 accessing the instrumentation variables, we use a thread-local 
data structure per thread, and merge data structures of the 
threads at the end of the program execution.

After the instrumentation step, the profiler receives the instrumented 
program with test input data, runs the program with each input, 
and collects tuples, each of which is program execution time and feature values 
for each input data. The profiler then sends the tuples 
of (execution time, feature values) to the prediction model
generator that performs feature selection and model creation.

\section{Prediction Modeling}
\label{sec:model}

We instrument and profile programs to collect many features from program execution runs for modeling the performance metrics of the programs.
However, we expect that a small but relevant set of features  
may explain the execution time well, and hence seek a compact model, 
i.e., a function of this small set of features, 
that accurately estimates the execution time of the program.
Among all the information, not all of them are expected to be 
useful for the model:
Some of them may have no variability across different
inputs, some have very weak or even no correlation 
to the execution time, and others are redundant to each other.
However, we do not know which features are useful, 
but would like to determine a small subset 
of features that is most relevant to predicting
the execution time, and are willing to sacrifice 
some of the small details in order to get the ``big picture".

To make the problem tractable, we constrain our models to the 
multivariate polynomial family. We expect that 
a good program should have polynomial execution time on 
some (combination of) features, and a polynomial model up 
to certain degree can approximate well any 
nonlinear model (due to Taylor expansion).
In addition, a compact polynomial model that predicts execution
time well can provide an easy-to-understand explanation on
what factors are important in determining the execution time 
of the program, and then give program developers intuitive feedback 
on the performance of the program.

In summary, what we need is an optimal strategy to produce a
(nonlinear) model on a small set of features from thousands of ones
collected blindly. We rely on machine learning techniques 
(specifically, sparse regression with multivariate polynomial basis)
to  automatically infer this small subset features and construct
a compact model to capture the dominant predictors of execution time.

\subsection{Background}
\paragraph{Least Square Regression}
Our feature instrumentation procedure outputs $n$ data 
samples as tuples of $\{y_i, \vecx_i\}_{i=1}^n$, where $y_i \in \R$ 
denotes the $i^{th}$ observation of execution time, 
and $\vecx_i$ denotes the $i^{th}$ observation of the vector of $m$ features.
We use regression techniques to model the relationship between 
$y$ and $\vecx$, which assumes that $y_i$'s are generated from
$y = f(\vecx, \beta) + \epsilon$,
where $\beta = [\beta_0, ~\beta_1, ~\beta_2, \ldots]$ 
is a vector of weights to be determined for the model,
and $\epsilon$ is the white noise.
Least square regression is a mathematical procedure for finding 
the best-fitting $f(\vecx, \beta)$ to a given set of responses 
$y_i$ by minimizing the sum of the squares of the 
residuals~\cite{ESL}, i.e.,
\begin{equation}
\label{eqn:ls}
\min_{\beta}\quad \sum_{i=1}^n\left(y_i-f(\vecx_i, \beta)\right)^2.
\end{equation}
If a linear function $f(\vecx, \beta)$ is used, we obtain 
linear least square regression, which can be easily extended to 
create nonlinear models by using
nonlinear (e.g., polynomial, spline, etc.) 
basis functions of features $\vecx$. 

\paragraph{Sparse Regression}
While widely used, least square regression has two major 
drawbacks: 1) When a large number of features exist, 
least squares tend to create complex models and  
overfit the data, resulting in inferior prediction accuracy.
2) It is usually hard to interpret the results, 
because it tends to create models involving many 
feature terms, if not all of them. 
This does not satisfy us since we have a lot of 
features but desire only a small subset of them to
contribute to the model.

Regression with best subset selection finds for 
each $k \in \{1, 2, \ldots, m\}$ the subset of size $k$
that gives smallest residual sum of squares. 
However, it is a discrete optimization 
and is known to be NP-hard~\cite{ESL}. 
In recent years a number of approaches based on model regularization 
have been proposed as efficient alternatives.
Their main idea is to add a regularization term to problem~\eqref{eqn:ls} to
control the complexity of the model, and make a tradeoff between the 
regression error and the number of features used in the model.
Among them, a widely used one is LASSO (Least Absolute Shrinkage and Selection Operator)~\cite{LASSO}, 
which uses quantity $\lambda \sum_{j=1}^m\abs{\beta_j}$ 
to penalize problem~\eqref{eqn:ls}. It effectively 
enforces many $\beta_j$'s to be 0, 
and selects a small subset 
of features (indexed by non-zero $\beta_j$'s) to build the model, 
which is usually compact and has better prediction accuracy 
than models created by ordinary least square regression~\cite{ESL}.
Parameter $\lambda$ controls the complexity of the model: 
as $\lambda$ grows larger, fewer features are selected by the model.

Being a convex optimization problem is the greatest advantage 
of the LASSO method, and there exist fast algorithms to solve the problem 
efficiently even with large-scale datasets~\cite{LARS, L1L2}. 
LASSO also has nice theoretical and empirical properties, 
and under suitable assumptions, it
can recover the true underlying model~\cite{LASSO, LASSO_OPT}.
In addition, LASSO can be easily extended to create  
nonlinear models (e.g., using polynomial basis functions of the features).

\subsection{Our Procedure}
We aim to use polynomial functions to model the execution time,
so that we can clearly see what kinds of nonlinear terms on 
which features are important to the execution time.
To capture nonlinear effects of and interactions between
multiple features, we expand the features 
$\vecx =[x_1~ x_2 \ldots ~x_k], k \leq m$ to 
all the terms in the expansion of the degree-$d$ polynomial  
$(1+x_1+\ldots+x_k)^d$, and use them to construct
a multivariate polynomial function $f(\vecx, \beta)$ for 
the regression. For example, using a degree-$2$ polynomial with 
feature vector $\vecx =[x_1~ x_2]$, we expand out $(1+x_1+x_2)^2$ 
to get terms $1, ~x_1, ~x_2, ~x_1^2, ~x_1x_2, ~x_2^2$, and use them
as basis functions to construct the following function for regression:
\begin{eqnarray*}
f(\vecx) = \beta_0 + \beta_1x_{1} + \beta_2x_{2} + \beta_3x_{1}^2 
+ \beta_4x_{1}x_{2} + \beta_5x_{2}^2 .
\end{eqnarray*}
Because we neither know which features are needed nor what kinds of 
nonlinear terms are necessary, an optimal but naive approach is
to expand the degree-$d$ multivariate polynomial with all 
$p$ features and use all the terms to construct the regression function.
However, this approach gives us ${m+d\choose d}$ terms, which is large 
when $m$ is on the order of thousands and even for small $d$, 
and will cause heavy burden on the computing of the regression model. 
Complete expansion on all features is not necessary, 
because many of them have little contribution to the execution time, 
and many of them are redundant to each other.

For efficient computation, we adopt a 3-step approach
for the feature selection and nonlinear model fitting:

\stitle{Step 1} Use the linear LASSO algorithm to filter out (many)
features that hardly contribute to the execution time. 
Although this step may be suboptimal (mainly due to the 
non-linearity in the true underlying model), it is cheap, fast and scalable, 
and is provably better than the traditional feature selection methods 
that consider individual features one by one.

\stitle{Step 2} Do degree-$d$ multivariate polynomial expansion on the
features selected in step (1), and use all the terms from the
expansion as the basis functions for the nonlinear model.

\stitle{Step 3} Use the LASSO method on the expanded features to pick out a 
subset of nonlinear terms to construct the model.

With these three steps, we have developed an efficient procedure 
to select a small set of nonlinear terms to construct a compact 
and intuitive model. Our experimental results 
in Section~\ref{sec:eval} show our method can construct
models to accurately predict execution time for a variety of 
applications.

% notations
\def\la{\langle}
\def\ra{\rangle}
\def\fun{\rightarrow}
\def\bigmid{\;\mid\;}
\def\M{\mathbb M}
\def\P{\mathbb P}
\def\V{\mathbb V}
\def\G{\mathbb G}
\def\F{\mathbb F}
\def\H{\mathbb H}
\def\E{\mathbb E}
\def\pset#1{\mathcal{P}(#1)}
\def\lc{{\tt lc}}

\section{Feature Evaluator Generation}
\label{sec:slicing}

In this section we explain our feature evaluator generation component.
To generate a feature evaluator, i.e., a small code snippet that computes 
the value of a feature, we use a program slicing algorithm.
Given a program and
a {\it slicing criterion}, which is a program variable $v$
at a program point $p$, {\it static slicing} \cite{Weiser81} computes a {\it slice}, which is an executable sub-program of the given program that yields the same value of $v$ at $p$ as the given program, on all inputs.
The goal of static slicing is to yield as small a sub-program as possible.
Figure~\ref{fig:ex-slice} shows a slicing example. The original code performs reading lines from a file, executes expensive computation on each line, and accumulates processed values. Suppose we want to extract the code part that affects the computation of the instrumented variable \texttt{mantis\_loop\_cnt1}. Ideally, the slicer should produce the sliced code, shown in Figure~\ref{fig:ex-slice} that captures only code that really affects the variable.
\begin{figure}[t]
{\small
\begin{verbatim}
// original code
int j;
while(line=readLine()) {
  ++mantis_loop_cnt1;
  j = j + expensive_processing(line);
}
// sliced code on the variable 
// mantis_loop_cnt1
while(line=readLine()) {
  ++mantis_loop_cnt1;
}
\end{verbatim}
}
\vspace{-0.2in}
\caption{A slicing example.}
\label{fig:ex-slice}
\end{figure}

At a high level, our slicer captures intra-procedural and inter-procedural
data dependencies and control dependencies of a slicing criterion. The 
produced slices must be executable since in our system the generated
sub-program will be executed online on the given input to obtain
the result of the slicing criterion. This is a requirement clearly 
different from most slicing research work motivated by debugging 
(e.g., \cite{Sridharan07}) whose goal is to
highlight as few statements as possible that will aid the
programmer debug a particular problem. Thus, they elide the constraint 
in the original slicing definition that the generated sub-program be executable. To achieve executability, we need to solve several engineering
issues related to Java language features.

Our slicing algorithm operates on expressions $e$, which may be of
one of four kinds: a local variable $v$, a static field (i.e., a global variable) $g$,
an abstract instance field $\la h,f \ra$ denoting instance field $f$ of
any object allocated at site $h$, or an abstract array element $h$,
denoting any element of any array object allocated at site $h$.
Abstractions of instance fields and array elements are required
because static analysis cannot refer to concrete object addresses.
Our slicing algorithm is not dependent upon the choice of abstraction,
however, can easily be modified to use abstractions besides
object allocation sites.\footnote{The choice of abstraction affects the precision and scalability of the algorithm, and we found object allocation sites to strike a good tradeoff.}

The slicer takes as input the given program and
the slicing criterion $c = \la e, p\ra$, which is an expression $e$
whose value is desired at program point $p$, and produces
as output a corresponding slice.
In our setting, $e$ is always a static field $g$ instrumented by
us (e.g., a loop counter), and $p$ is always the exit of a method of the 
program (e.g., the program's main method.)

Our slicing algorithm is based on two algorithms (one from Horwitz, Reps, and Sagiv~\cite{Horwitz88} and one from Reps, Horwitz, Sagiv, and Rosay~\cite{Reps1994}). We summarize four steps of the algorithm. First, for each method,
we construct a Program Dependence Graph (PDG). Then, for the entire program,
we construct a System Dependence Graph (SDG), which is a set of PDGs where additional edges are created to capture interprocedural dependencies. 
We augment the SDG with summary edges by running the interprocedural data flow analysis algorithm in~\cite{Reps95} to solve context sensitivity problems. 
Finally, we run a 2-pass reachability algorithm on the augmented SDG. We explain individual steps more in detail below.

\paragraph{PDG and SDG}
Our slicing algorithm operates on Joeq~\cite{joeq} quad code, an intermediate
representation format based on registers. The vertices of a PDG represent
quad code instructions (e.g., statements and predicates). The edges of a PDG
represent data flow and control dependencies. An SDG also includes
inter-procedural dependencies. A method call creates a call vertex and a set
of actual-in and actual-out vertices. Each parameter of a method call creates
an actual-in vertex, and a return value creates an actual-out vertex.
A method entry creates an entry vertex and a set of formal-in and formal-out
vertices, which correspond to arguments and a return value respectively.
A call edge is created to connect a call vertex of an call site to an entry
vertex of the matching method. In addition, a linkage-entry edge is created
from an actual-in vertex to a corresponding formal-in vertex, and a
linkage-exit edge is created to link a formal-out vertex to an actual-out
vertex.

\paragraph{Augmented SDG}
An SDG does not capture context-sensitivity of method calls. Therefore, if a
call site of a method is included in a slice, other call sites of the same
method may be included even though they do not affect a slicing criterion. 
To remedy this problem, we build an augmented SDG by adding {\it summary edges} to an SDG.
A summary edge connects an actual-in vertex to an actual-out vertex and summarizes
the effect of the actual-in on the actual-out of a method call.
To create summary edges, we use the backward RHS algorithm~\cite{Reps95}. It propagates
from formal-out vertices of the method based on data- and control- dependencies
to calculate path edges of the method. A path edge is of the form $\la p, e_1 \ra \fun
 \la p_{formal-out}, e_2 \ra$ meaning that $e_1$ at program point $p$ affects $e_2$ at $p_{formal-out}$ :
it always ends with a formal-out of the method and is created within the method. 
Therefore, when there exists a path edge from a formal-in vertex to a formal-out vertex of
a method, a summary edge is created connecting corresponding actual-in vertex and 
actual-out vertex of a call site to the method.

\paragraph{2-pass algorithm}
To identify statements to include in a slice, we run a 2-pass reachability
algorithm on an augmented SDG. The first pass starts from the program point
of a given slicing criterion and goes backwards along all the edges in the
augmented SDG but {\it not} along linkage-exit edges. As a result, when
encountering a call site, the first pass does not go into the method body
but uses summary edges of the call site.
The second pass starts from all actual-out vertices visited in the first pass
and traverses backwards using all the edges but {\it not} using linkage-entry
and call edges. This pass covers all methods that correspond to call sites
identified in the first pass. In addition, it may find more call sites while
traversing the body of the methods and use summary edges; this process adds
additional actual-out vertices of the summary edges and the pass goes into
the methods associated with the actual-outs. 

\paragraph{Slicing made practical}
A set of program statements identified by the described algorithm may not
meet Java language requirements. This problem needs to be resolved to create
executable slices. We list a few of the engineering issues we addressed for that.
First, we need to handle accesses to static fields and heap locations (instance
fields and array elements). Therefore, when building an SDG, we identify all
such accesses in a method and create formal-in vertices for those read and
formal-out for those written along with corresponding actual-in and actual-out vertices.
Second, there may be uninitialized parameters if they are not included
in a slice. We opt to keep method signatures, hence we initialize them with
default values. Third, there are methods not reachable from a main method but
rather called from JVM directly (e.g., class initializers). These methods will not
be included in a slice by the algorithm but still may affect the slicing criterion.
Therefore, we do not slice out such code. Fourth, when a new object creation is
in a slice, a corresponding constructor invocation may not.
To address this, we create a control dependency between object creations and
corresponding constructor invocations to ensure that they are also in the
slice. Fifth, a constructor of a class except the Object class must include
a call to a constructor of its parent class.
Hence we include such calls when they are
missing in a slice. Sixth, the first parameter of an instance method call is
a reference to the associated object. Therefore if such a call site is in a slice,
the first parameter has to be in the slice too and we ensure this.

\paragraph{Final step}
Previous steps we described so far generate a slice of Joeq quad code.
To generate the final Java byte code we can execute, we translate 
the Joeq quad code to Jasmin~\cite{jasmin}
assembly code and use the Jasmin assembler to generate Java byte code. We take a
simple approach that translates each quad instruction to a corresponding set
of byte codes. During the process, since we do not have complete information
on ordering between basic blocks, we add an explicit {\it goto} instruction at
the end of each basic block. However this may lead to a cycle if a conditional
branch in a loop is sliced out and replaced by goto. We ensure that no cycle is
created by performing a DFS-like search and choosing a successor as 
a target of the
goto instruction only if it can reach the exit of the method. Another special
case is JSR instruction that pushes the address of the next immediate opcode
into an operand stack as its return address. However the next instruction may
not be the same as one in the original program. Hence we add an extra goto with
an appropriate target after the JSR operation.
Our current translator is not optimized; 
we plan to optimize the use of stacks if needed in the future.

\paragraph{Discussion}
There are static and dynamic program slicing algorithms. They have tradeoffs between input coverage and slice compactness. Static slicing works for all inputs, but it may produce a bigger slice than dynamic slicing. Dynamic slicing includes only code that is actually executed for given inputs, but it does not cover all inputs. As a starting point, we chose static slicing as it guarantees to work for all inputs, but in the future we plan to explore dynamic slicing or hybrid slicing that combines static slicing and dynamic slicing if we need to improve slice compactness. In this paper, we tested our static slicing algorithms with simple programs, and we plan to evaluate the scalability of our algorithms with complicated programs.

\section{Evaluation}
\label{sec:eval}
We have implemented Mantis that works with Java programs by extending existing machine learning and program analysis tools.
We built the feature instrumentor atop Eclipse JDT AST libraries. JDT is a toolkit that provides APIs to access and manipulate Java source code. We add visitors to ASTs to add instrumentation code. The basic instrumentation variables are thread local variables. The instrumentor also introduces static global variables that summarize feature values across threads and are used for slicing. We implemented our modeling procedure in Matlab. Finally, we extended JChord~\cite{jchord}, a static and dynamic Java program analysis tool, to implement our program slicing algorithm in Java and Datalog. In the current version, we have to patch models for native library functions manually to track dependencies inside native library functions. The JChord slicer produces Joeq quadcode slices. To create final executable bytecode slices, we implemented a translator from quadcode to bytecode using Jasmin~\cite{jasmin}. 

To evaluate our system, we choose two applications,
Lucene Search~\cite{mahout} and ImageJ~\cite{ImageJ},
that involve intensive computation.
We evaluate the prediction accuracy of our system in terms of 
prediction error (i.e., prediction accuracy = 1 - prediction error) 
and compare it with blackbox approaches. Prediction error is computed using
the equation:
\[
\textrm{prediction error} = \frac{ \abs{\textrm{predicted time - actual time}}}{\textrm{actual time}}.
\]
We show the sensitivity to the size of training data and the regularization parameter $\lambda$, and the results of prediction.
Traditional blackbox approaches fail to predict execution time with low prediction error, but our system can construct an online predictor that can predict execution time accurately.
Note that, in presenting the results of our system, we use only features that can be compuated cheaply by iterating over feature selection and program slicing for rejecting expensive features.

\begin{figure}[t]
\centering
\includegraphics[width = 0.7\columnwidth]
{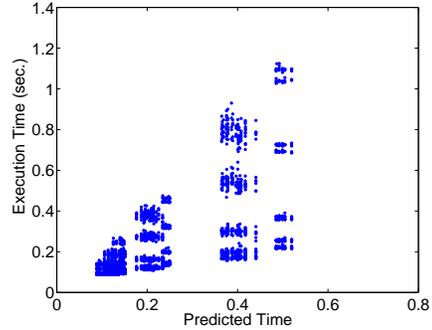}\\
\caption{For Lucene, the blackbox approach fails to predict execution 
time accurately.} 
\label{fig:lucene_strawman}
\end{figure}

\begin{figure}[t]
\centering
\includegraphics[width = 0.7\columnwidth]
{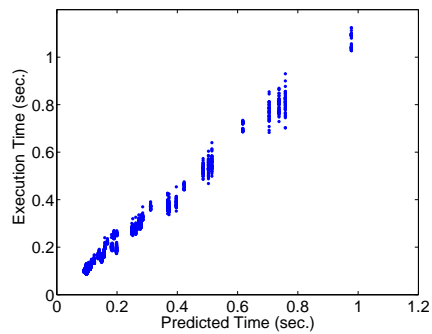}\\
\caption{
Predicted time vs. execution time of the Lucene search application.
Using 10\% of Lucene search data for training, 
our system can predict execution time with less than 7\% prediction 
error.
}
\label{fig:lusearch_features}
\end{figure}

\subsection{Prediction Results for Lucene}
\label{sec:eval-lucene}

After profiling our Lucene search application with various text input queries
over a corpus of the works of Shakespeare and the King James Bible,
we obtain a dataset with 3840 samples, each
of which consists of 1 execution time, 9 loop features, 29 branch features,
and 90 variable features from 18 variables (we record 5 versions of
values for each variable). So we obtain a dataset with 1 column of
execution time and 126 columns of features, subset of which would
hypothetically explain the execution time well. 
In the prediction modeling process,
we normalize each column of values into range [0,1],
and randomly partition data (row) samples into training set and testing set.

\begin{figure*}[t]
\vskip -0.1in
\begin{center}
\begin{tabular}{cc}
\includegraphics[width = 0.8\columnwidth]
{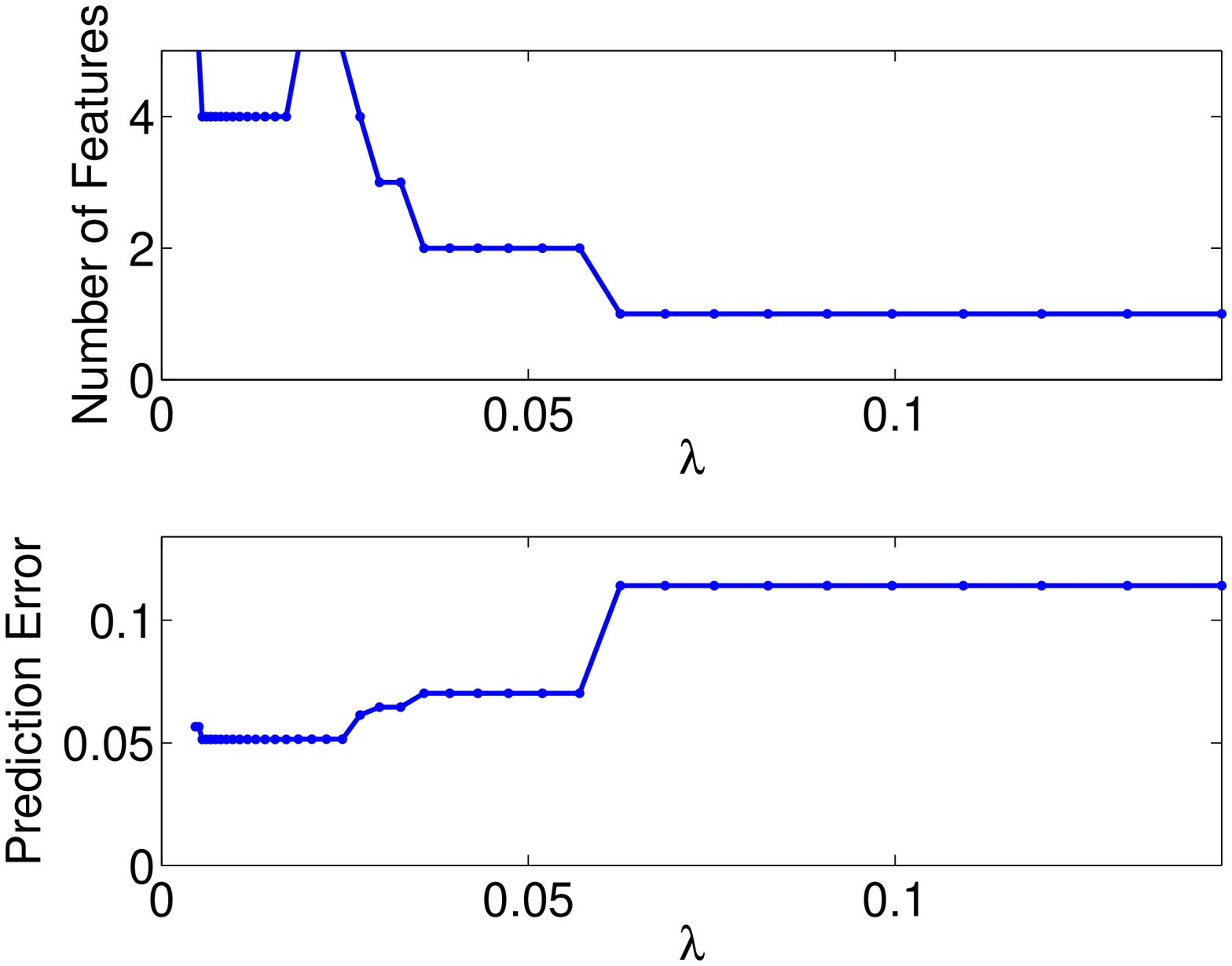} &
\includegraphics[width = 0.8\columnwidth]
{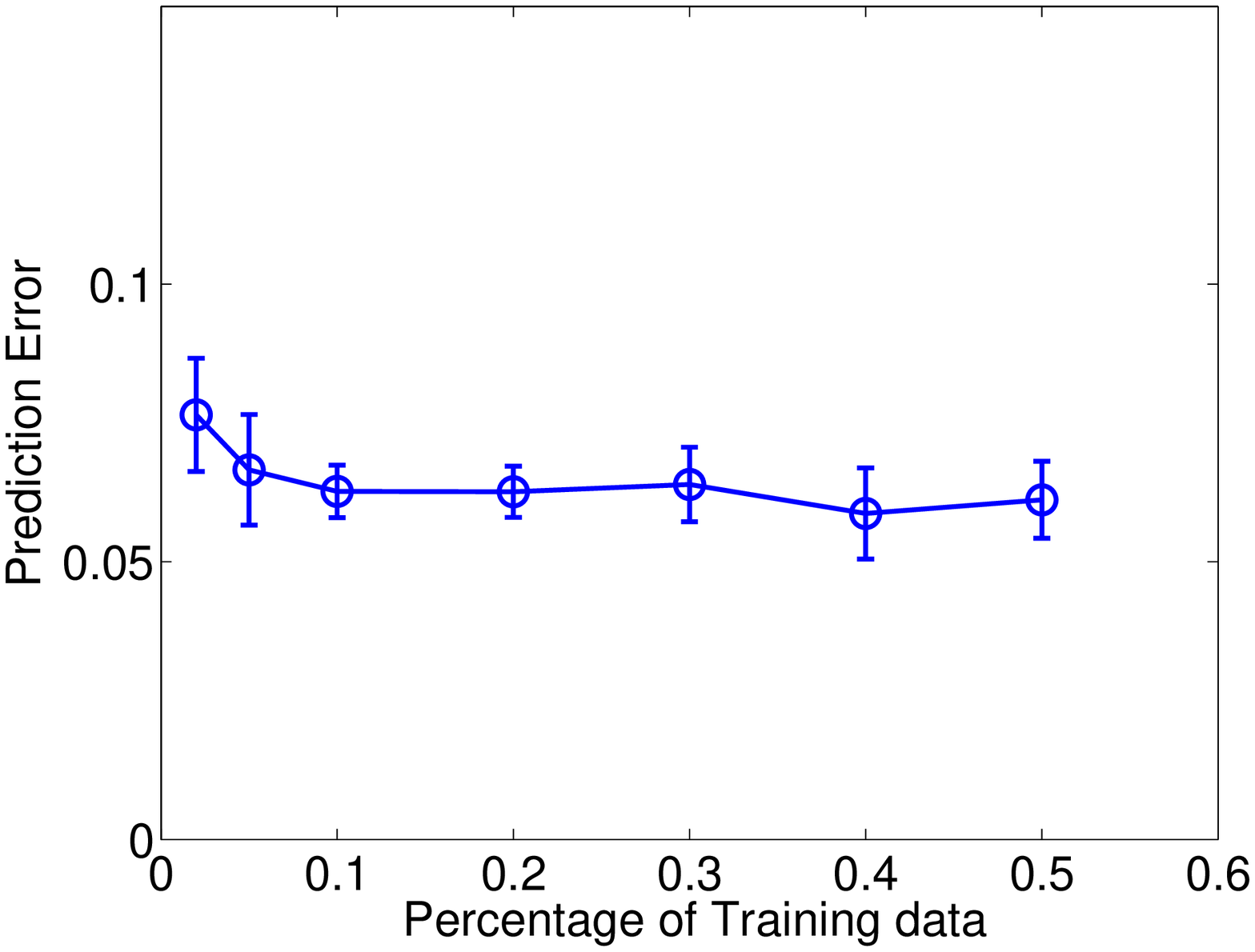}
\\ {(a)} & {(b)}
\vspace{-0.1in}
\end{tabular}
\caption{{ Using Lucene search data, we show in (a) that our 
approach is insensitive to parameter $\lambda$:
there are a range of $\lambda \in (0, 0.07]$ that result
in similar models for accurate prediction, and show in (b) that our
approach is insensitive to the size of training data:
even using 5\% or less of data for training, our system can create
models achieving accurate prediction for execution time.}}
\label{fig:lusearch_sensitivity}
\end{center}
\vskip -0.2in
\end{figure*}

We first evaluate a blackbox approach for predicting
execution time. We choose one that 
can construct \textit{compact} nonlinear models using 
the command line arguments as
features (instead of using the feature data from our profiler), which
consist of the following ones:
{\tt raw, threads, totalqueries, hitsperpage, repeat},
denoted by $x_1$, $x_2$, $x_3$, $x_4$ and $x_5$, respectively.
We build models using either ordinary least-square regression
or LASSO with a function using all terms in the
expansion of $(1+x_1+x_2+x_3+x_4+x_5)^3$. However, in either case
we consistently see more than 38\% prediction error, even when we:
1) (randomly) sample different portions of data for training,
2) vary the size of training data (for model regression) from 10\%
to 40\%,
3) use polynomial functions with order higher than 3, and
4) use different subsets of the 5 features.
Figure~\ref{fig:lucene_strawman} shows predicted execution time vs. 
actual execution time.
As you can see, this blackbox approach derived from command line features
fails to model and predict execution time accurately.

From our detailed analysis, 
two features that have the largest correlation with the
execution time are 
feature-1, {\tt totalqueries}, which has a fair amount of
correlation with execution time, and feature-2, {\tt  thread},
and the remaining features are poorly correlated with execution time.
Despite some correlation in feature-1
the model derived from command line features
are not enough for predicting execution time accurately:
for each value of the predicted time (on $x$-axis),
there are dramatically different actual execution times
(on $y$-axis) correspond to (an ideal prediction is a $45$
degree line pass through the origin).
This result indicates that
some other factors that are not captured by the features should
contribute to the execution time.
On the contrary, as shown in the following, our system can automatically
select higher quality features from the program, and construct nonlinear
models to predict execution time accurately.

To evaluate our system, we start with its sensitivity to $\lambda$,
the parameter for trading off the prediction error with the number of
selected features. We use 10\% of data for training (both feature
selection and model fitting), and trace a variety of $\lambda$ values
(which may result in different subset of selected features,
thus different models) using an efficient algorithm
proposed in~\cite{LARS}. We show the result in
Figure~\ref{fig:lusearch_sensitivity} (a).
To our surprise, we see that our method
is able to select 2-4 features (out of 126 in total) that are enough
to build a nonlinear model to predict the execution time within 7\% error.
As expected, starting from a very small $\lambda$ value and increasing it,
our method selects decreasing number of features (from 4 down to 1),
and consequently results in models with decreasing prediction power.
We clearly see that there is a range of $\lambda$ values (e.g., (0, 0.07])
that enable our method to select the right set of features and
build models for accurate prediction.
Repeating the experiment with different (random) training samples,
and with 20\%, 30\% and 40\% of data for training, we see
very similar behavior. We conclude that our method
is insensitive to the parameter $\lambda$, and setting
$\lambda \leq 0.07$ allows us to select right features, and construct
compact and accurate models for predicting execution time.

Fixing $\lambda=0.03$\footnote{An optimal $\lambda$ can be determined by
a cross-validation approach, e.g., further partitioning the training data
into two sets, one for feature selection and model regression, and another
for testing the model. An optimal $\lambda$ is the one giving the
smallest testing error (on the part of training data selected for testing).},
we study the sensitivity of our method to the
size of training data, and plot the result in
Figure~\ref{fig:lusearch_sensitivity} (b). We see that with different sizes
of training data, prediction errors of the constructed models are
fairy stable, and even using 5\% or less training data, our method
is able to produce accurate models for predicting execution time.

To reveal more details of the model, we use $\lambda=0.03$ and
10\% of data for training, to investigate which features are
selected and what kinds of models are constructed.
We find that our algorithm usually selects 3-4 features (depending on
which subset of data are sampled for training) and constructs
a model with a prediction error around 6.1\%. 
Figure~\ref{fig:lusearch_features}
shows predicted execution time vs. actual execution time of our system.
In one instance of modeling, the following 4 features are selected:
1) loop feature $l_2$ related to a while loop for reading
keywords from the query file,
2) variable feature $v_3$ related to {\tt totalQueries},
3) variable feature $v_5$ related to {\tt hitsPerPage}, and
4) variable feature $v_9$ related to how many query
processors to create per thread.
Among them, features $l_2$ and $v_9$ have the largest weights
(indicating they are the most important) and persistently
appear when sampling different portions of the data for training.
With just these two features, we do a LASSO sparse regression with
all basis functions of features in the expansion of $(1+l_2+v_9)^3$.
Interestingly, we are able to construct the following a nonlinear model
\begin{eqnarray*}
f(l_2, v_9) =  0.1 + 0.52 l_2 + 0.09 v_9 - 0.69 l_2^2 - 0.07v_9^2 + \\
+ 1.16 l_2^3 + 0.13 l_1v_9^2,
\end{eqnarray*}
which can predict execution time with error 6.7\% (indicating the
rest two selected features $v_3$ and $v_5$ only contribute to
less than 1\% of the prediction accuracy).

\begin{figure}[t]
\centering
\includegraphics[width = 0.7\columnwidth]{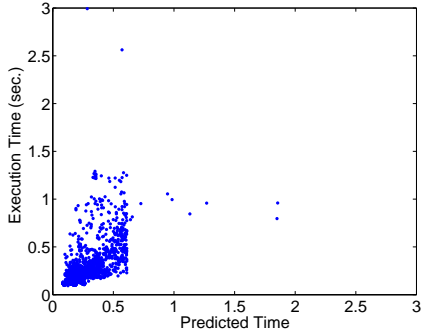}
\caption{Predicted time vs. execution time of the ImageJ application for a blackbox approach.}
\label{fig:imageJ_strawman}
\end{figure}

\begin{figure}[t]
\centering
\includegraphics[width = 0.7\columnwidth]{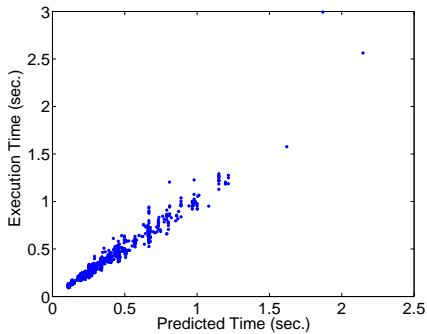}
\caption{Predicted time vs. execution time of the ImageJ application. Using 10\% of ImageJ data for training, our system can predict execution time with 5.5\% prediction error.}
\label{fig:imageJ_features}
\end{figure}

\subsection{Prediction Results for ImageJ}
\label{sec:eval-imagej}
ImageJ~\cite{ImageJ} is a public domain Java image processing and
analysis program. It provides a variety tools for displaying,
editing, analyzing, and processing images in many formats.
We test a dozen of tools of ImageJ, including {\tt Smooth, Find Edges, FFT,
Find Maxima,} etc. We choose to profile and predict the execution time
of {\tt Find Maxima}, because it exhibits high variance in execution
time when processing different images (even with similar size), making
it a challenging task to model the execution time.

To profile the {\tt Find Maxima}, we use 3045 images
from popular vision corpus of Caltech 101~\cite{Caltech101}, Event
Dataset~\cite{EventDataset} and PASCAL challenge 2008 dataset~\cite{PASCAL}.
The images vary a lot in size and resolution, and have content in different
scenes (e.g., in the office, on the street, in the natural environment, etc)
and with different object categories (e.g., plan, car, bird, building, etc).
After the profiling, we obtain a dataset
with 3045  samples, each
of which consists of one execution time, 291 loop features, 2935 branch features,
and 2290 variable features from 458 variables (we record five versions of
values for each variable). So we obtain a dataset with one column of
execution time and 5516 columns of features. After removing constant and
redundant columns, we obtain 182 useful features, (small) subset of which would
likely explain the execution time well. In the experiments,
we normalize each column of values into range [0,1],
and randomly partition the data into training set and testing set.

For a blackbox approach, many methods can be used. We consider one with
the execution time as a nonlinear function of a simple input
parameter -- the image size. We start with a degree 3 polynomial
function of the image size $x$, and obtain the following prediction
model using 20\% of data for training
\begin{equation*}
f(x) = 0.1 + 2.18x -8.77 x^2 + 36.6x^3.
\end{equation*}
Predicting on the remaining 80\% test data, we consistently see
more 35\% of prediction error regardless which subset of data are
sampled as a training set (may result in slightly different models).
We see similar results when we increase
the degree of the polynomial and the percentage of training data.

We plot the execution time against the predicted execution time obtained from the model in Figure~\ref{fig:imageJ_strawman}. We clearly
see that image size alone is not enough for predicting execution time
regardless of whatever model may come out: the image size is poorly
correlated with the execution time, and for each value of the image size
or the predicted execution time, there are dramatically different actual
execution times corresponding, indicating that some other factors
should contribute to the execution time.

On the contrary, our system can automatically select two high-quality 
features from thousands of automatically instrumented features,
and construct a model to predict execution time accurately.
Of the two selected features,
%(variables 97 and 200)
one is the variable feature related to the width of region of interest
(denoted by $w$); the other is height of the image (denoted by $h$).
Although highly correlated to the execution time, neither
a single feature (even with nonlinear model), nor the linear
combination of both selected features can predict execution time very well.
Instead, using 10\% of data for training with $\lambda = 0.03$,
we do a LASSO sparse regression using all (nonlinear) terms in the
expansion of $(1+w+h)^3$, and obtain the following model
\begin{eqnarray*}
f(w,h) = 0.1 + 0.08w + 0.07h + 0.33wh + 0.02h^2,
\end{eqnarray*}
which can predict the execution time accurately (around 5.5\%
prediction error), as shown in Figure~\ref{fig:imageJ_features}.

\begin{figure*}[t]
\vskip -0.1in
\begin{center}
\begin{tabular}{cc}
\includegraphics[width = 0.8\columnwidth]
{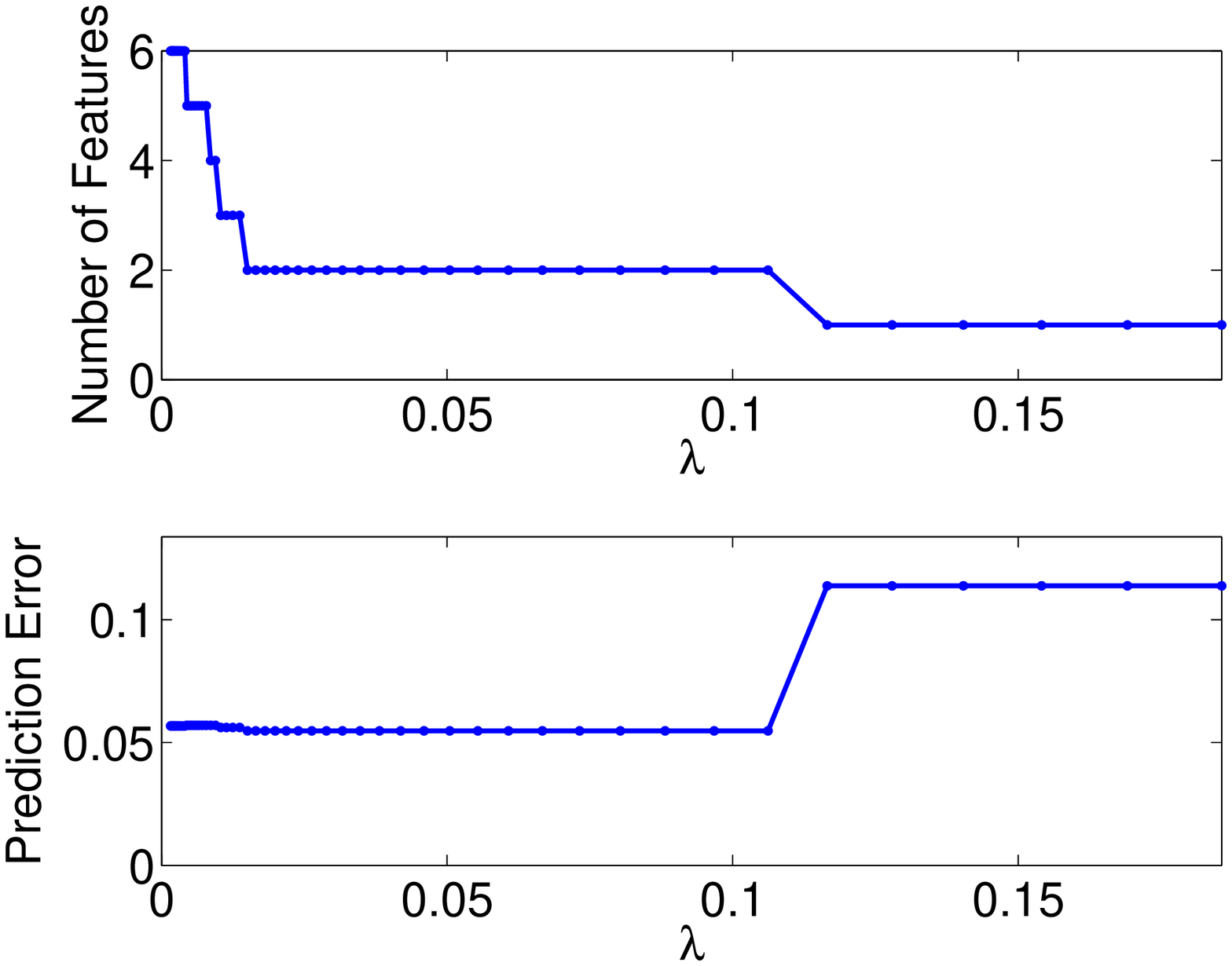} &
\includegraphics[width = 0.8\columnwidth]
{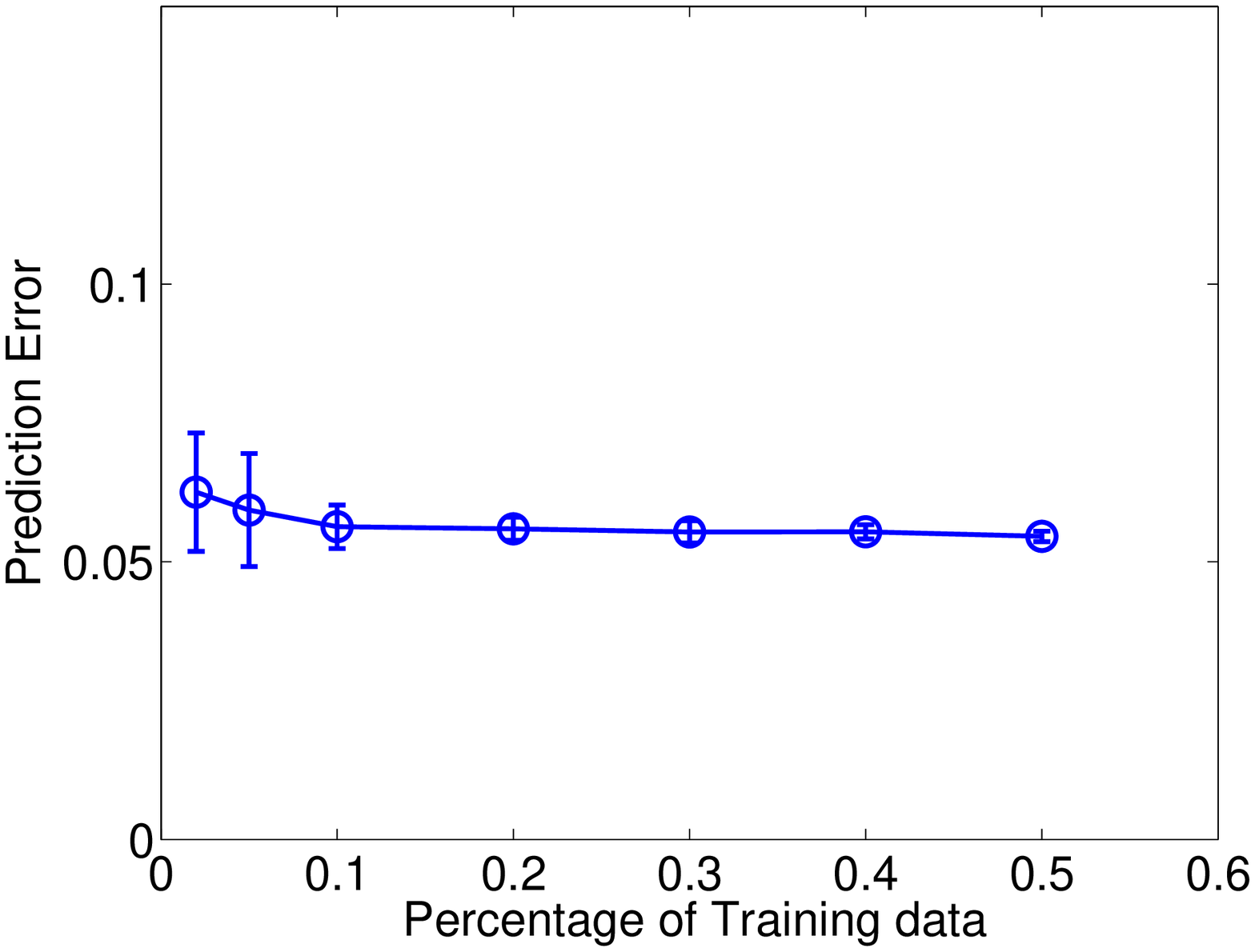}
\\ {(a)} & {(b)}
\vspace{-0.1in}
\end{tabular}
\caption{{Using ImageJ data, we show in (a) that our approach is 
insensitive to $\lambda$:
there are a range of $\lambda \in (0, 0.11]$
that result in similar  models for accurate prediction,
and show in (b) that our approach is insensitive to the size of training data:
even using 5\% or less of data for training, our system can 
create models achieving accurate prediction. 
}}
\label{fig:imageJ_sensitivity}
\end{center}
\vskip -0.2in
\end{figure*}

We also study the sensitivity of our method to $\lambda$
using 10\% of data for training (both feature
selection and model fitting), and show the result in
Figure~\ref{fig:imageJ_sensitivity} (a). Again, we see that our
method is insensitive to $\lambda$, and there are a wide range
of $\lambda$ values (e.g., $(0, 0.11]$) that allow us to select
right features, and construct compact and accurate models
for predicting execution time.

Fixing $\lambda=0.03$, we study the sensitivity of our method to the
size of training data, and plot the result in
Figure~\ref{fig:lusearch_sensitivity} (b). Again, we see that with
different sizes of training data, prediction errors of the
constructed models are fairy stable, and even using 5\% or less
training data, our method is able to produce accurate models for
predicting execution time.

\subsection{Benefit of Slicing}
In this section, we evaluate the benefit of slicing. We explain how slicing can help choose features that are not expensive to compute and evaluate the execution times of feature evaluators of selected features computed manually.

\begin{table}
\begin{tabular}{l|l|l}
Step & Features selected by  & Rejected \\
     & the model generator & features \\\hline\hline
1 &  loop features $l_3$ and $l_7$ & loop feature\\
 &  variable features $v_3$, $v_5$, and $v_9$ & $l_3$\\\hline
2 &  loop feature $l_2$ & NONE \\
 &  variable features $v_3$, $v_5$, and $v_9$ &  \\
\end{tabular}
\label{tab:feedbackloop}
\caption{Iterative procedure of model selection considering the cost of computing feature values. In the example, loop feature $l_3$ is related to a for loop printing search results, and loop feature $l_7$ is related to a while loop counting how many times queries are executed. We explained other features in Section~\ref{sec:eval-lucene}.
Computing $l_3$ requires computing the most of the program (i.e., doing actual lookups of indices), thus it is rejected in step 1. This iterative procedure stops at step 2 since all selected features are quickly computable.}
\end{table}

We look at the details on how the Lucene search application chose final features we presented in Section~\ref{sec:eval-lucene}. Table~\ref{tab:feedbackloop} shows the steps taken by the model generator due to the feedback from the slicer. In each step, we show the features selected by the model generator, and the features rejected by the slicer among the features passed from the model generator. For this application, at step 2, the slicer can compute slices that can quickly compute all the features needed by the model, thus it accepts the selected features and the feedback loop from the slicer to the model generator ends.

\begin{figure}[t]
\vskip -0.1in
\begin{center}
\includegraphics[width=0.8\columnwidth]{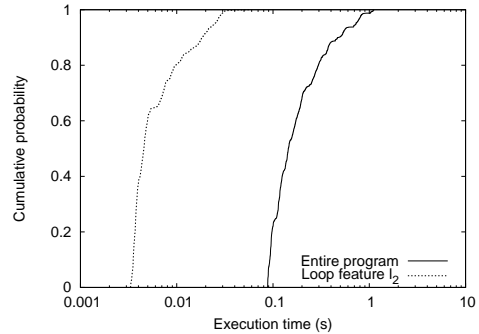}
\vspace{-0.1in}
\caption{CDF of execution time of the entire Lucene search program and time to execute the slice that computes loop feature $l_2$.}
\label{fig:slice-time-cdf}
\end{center}
\vskip -0.2in
\end{figure}

Figure~\ref{fig:slice-time-cdf} shows the cumulative distribution function (CDF) of execution time of the entire Lucene search program and that of the slice to compute loop feature $l_2$. We show only $l_2$ because it is the most expensive selected feature to compute since the slice goes through files to count keywords. The other variable features are derived by arithmetic operations and assignments of values from inputs. Computing $l_2$ takes $3-4\%$ of the entire program execution time, thus the prediction model can compute an estimate of execution time with low overhead.

\section{Related Work}
\label{sec:related}
Prediction has been explored in multiple different contexts --- database, cluster and cloud, networking, and program complexity modeling. In this paper, we presented a new performance prediction framework for generic programs by combining programming language and machine learning techniques. As far as we know, our work is the first to explore program analysis to extract features, employ machine learning to create accurate models with selected features, and use program slicing to automatically produce code snippets that compute feature values for prediction. 

In the database, researchers explored machine learning algorithms to predict database query execution time. Gupta, Mehta, and Dayal~\cite{Gupta2008} used
a variant of decision trees to predict time ranges of data warehouse 
queries. Ganapathi et al.~\cite{Ganapathi2009} used KCCA to predict time
and resource consumption of database queries (number of I/Os and messages) 
using the statistics of query texts and query plans (e.g., instance count
for each possible database operator).

In resource allocation and provisioning for cluster and cloud applications, 
research has been done to forecast how much resource is required to support given workload to meet service level agreements~\cite{Bodik2009a, Bodik2009b}, or how long it takes to complete a candidate assignment~\cite{Shivam2006}. The models used resource consumption or workload size for prediction. Xu et al. used console logs --- coarse-grained program status reports --- to detect anomalies in the Hadoop Distributed File System and the Darkstar game~\cite{Xu2009}. 

For load shedding in network monitoring applications, 
Barlet-Ros et al.~\cite{Barlet-Ros2007} used a simple linear regression model
of features from five packet header tuples, number of bytes, number of packets to predict CPU resource usage. This work is specific to packet processing applications. In contrast, our framework is applicable to generic programs.

In the networking context, multiple projects addressed the problems of predicting response time changes for what-if scenarios. 
WebProphet~\cite{Li2010} predicts the impact of certain optimizations
of web services before deploying them by extracting web object dependencies with injected delays and simulating web page loading processes with web object dependency graphs.
WISE~\cite{Tariq2008} predicts the effects of configuration or
deployment changes in content distribution networks by modeling the network dependency structure to response-time distribution.
Link Gradients~\cite{Chen2009} predicts the impact of network latency in multi-tier systems by doing delay injection and performing spectral analysis.

There have been studies on using information from execution traces for modeling computational complexity~\cite{Goldsmith2007}, simulating hardware platforms efficiently~\cite{Sherwood2001,Sherwood2002}, and finding bugs cooperatively~\cite{Liblit2005}.
In contrast to these, Mantis focuses on creating a model for predicting program execution time by computing feature values online with slices quickly for new inputs. 

Trendprof~\cite{Goldsmith2007} models asymptotic computational complexity by measuring empirical computational complexity. It computes a model that estimates the performance of a program by modeling basic block execution frequency in terms of user-specified features (e.g., input size) and summarizing the program with clusters of basic blocks.

SimPoint~\cite{Sherwood2001,Sherwood2002} finds a subset of execution instruction traces of program for an input for efficient hardware platform simulation because simulating hardware for the entire program execution takes too long time. It instruments basic block vectors in each fixed interval and uses a clustering algorithm to extract a representative subset of traces from clusters that approximates low-level hardware metrics such as instructions per cycle, percent RUU occupancy, cache miss rate, branch prediction miss rate, and address prediction miss rate~\cite{Sherwood2002}. 

Cooperative bug isolation (CBI)~\cite{Liblit2005}
used three predicates --- branches with four values (always true, always false, sometimes true and sometimes false, unreachable in the run), comparisons between all pairs of integer-valued variables and constants in the program, and comparisons between integer valued return results of functions
and 0. CBI aims  to find which predicates are correlated with crashes to find bugs, and uses sampling of predicates to lower runtime overhead since the executed runs are collected from end users of the program.

\section{Conclusion}
\label{sec:conc}
% impact of our work
In this paper, we presented Mantis, a new prediction framework that extracts program features using code analysis, models performance with these features using sparse regression, and generates code snippets that compute the feature values. We take a first step towards building such a framework. Our prototype evaluation shows that Mantis can predict execution time with more than 93\% accuracy for the applications we tested, a search engine and an image processing application, which cannot be achieved with models without program features. In the future, we plan to evaluate our system with various complicated applications in terms of accuracy, applicability, and scalability.

% future work
Our new approach to prediction presents several exciting research directions we want to explore.
First, we want to extend our model to include environment and to explore more sophisticated features (e.g., feature values that depend on calling contexts) and more sophisticated slicing algorithms (e.g., algorithms based on dynamic control flow graphs). Second, we would like to build our framework for C/C++ languages with LLVM~\cite{llvm} since the current prototype works with Java programs. 
Third, we want to further extend our framework to apply to networked systems running on multiple nodes. Our work in this paper addressed single-machine program execution. Finally, we also would like to apply the tool to performance debugging (e.g., a tool that generates test cases for performance debugging similar to KLEE~\cite{Cadar2008} that generates test cases for correctness).

%{\footnotesize
\bibliographystyle{abbrv}
\bibliography{mantis}
%}

\end{document}